# THE APPLICATION OF A DATA MINING FRAMEWORK TO ENERGY USAGE PROFILING IN DOMESTIC RESIDENCES USING UK DATA

## Ian Dent[1], Uwe Aickelin[2], Tom Rodden[3]

[1,2]*Intelligent Modelling and Analysis Group, University of Nottingham*
[1,3]*Horizon Digital Economy Research Institute, University of Nottingham*

**Abstract:** This paper describes a method for defining representative load profiles for domestic electricity users in the UK. It considers bottom up and clustering methods and then details the research plans for implementing and improving existing framework approaches based on the overall usage profile. The work focuses on adapting and applying analysis framework approaches to UK energy data in order to determine the effectiveness of creating a few (single figures) archetypical users with the intention of improving on the current methods of determining usage profiles. The work is currently in progress and the paper details initial results using data collected in Milton Keynes around 1990. Various possible enhancements to the work are considered including a split based on temperature to reflect the varying UK weather conditions.

**Keywords:** *Load profiles, clustering, data mining, electricity markets, demand-side*

## 1 Introduction

The UK electricity market is moving towards having the ability to provide more targeted and more complicated tariff offers for customers in order to provide many benefits including maximising the efficiency of the supply process. (DECC, 2009) shows that the provision of Smart Meters will allow greatly increased analysis of a customer's electricity usage and provide the ability to make customised offers on pricing and availability in order to change customer behaviour (for example, to minimise usage during peak periods) or to increase efficiencies in the electricity supply chain in meeting the predicted demand. (Ofgem, 2010).

The identification of typical electrical usage patterns within households is necessary as a starting point for:

---

[1] Ian Dent, ird@cs.nott.ac.uk (corresponding author)
[2] Professor Uwe Aickelin, uwe.aickelin@nottingham.ac.uk
[3] Professor Tom Rodden, tar@cs.nott.ac.uk

- Assessing the impact of any initiatives to reduce overall energy usage in order to discover the amount of overall reduction which occurs during different times of the day.
- Allowing accurate aggregation to provide a pattern of total demand to be met by supply side generation and transmission.

The work in progress forms part of a "demand side maximisation" project and will focus on identifying typical usage profiles for households and then clustering them into a few archetypical profiles with similar kinds of customers grouped together. Differences between an individual household profile and that of others within the same group can be used to suggest energy usage behaviour changes to reduce overall electricity usage or to improve electrical efficiencies possibly by time shifting particular appliance usage. In addition, particular groups (for example, large users during peak times) can be identified for targeting for reduction initiatives.

The work focuses on the daily variances of a household's electricity usage averaged over appropriate weekly, monthly or seasonal periods. Investigation of consumers' usage of electricity in order to determine similarities between types of consumers requires that the day's usage pattern is summarised in some way such that it can be compared with others. The "shape" of the usage pattern (e.g. little night usage, peak around breakfast, little usage during the day and then a peak during the evening period) needs to be determined.

The main purpose of the work is to investigate whether it is possible, using a clustering approach, to find representative daily usage profiles for a population of domestic residences in the UK and to investigate separation between the discovered typical profiles.

## 2  Profiling Approaches

A "bottom-up" approach can be taken to generating a representative load profile where the load generated by individual appliances is aggregated to give the overall household usage pattern. This is explored by (Capasso et al., 1994) and (Paatero & Lund, 2006) with more recent work done in the UK by (Richardson et al., 2010). With a bottom up model, each electrical load within the household can be documented and new, possible households with varying mixes of appliances can be created. An overall view of total load can then be calculated by aggregating the individual appliances and electrical loads over the household and then over many households. This approach is suitable for investigation using a simulation approach as the numbers and usage of each type of appliance and household type can be modified and then run in a simulation in order to determine the overall electricity usage under a given set of input parameters describing the appliance types and usage times.

A "top-down" approach can be taken by analysing the overall electricity usage of the household without regard to usage of particular appliances or electrical circuits. The overall shape of the daily usage pattern can be analysed using data mining tools and measures of similarity to other households or other times (previous or future days) for the same household can be calculated.

(Ramos & Vale, 2008) present a methodology to characterise new electricity customers into similar groups based on their load profiles (daily usage patterns). In order to reduce the amount of data, the load profiles for a given customer for a given type of day (weekend, weekday, or holiday) are combined to give a few representative profiles for that customer. For their case study in the paper, two types of day were considered (weekend and weekday) so that each of the 208 consumers are allocated 2 representative load profiles..

(Electricity Association, 1997) identifies a process for defining the details of 8 different standard usage profiles for the UK which are listed in Table 1.

**Table 1:** Profile Classes

| Profile class | Description |
|---|---|
| Class 1 | Domestic Unrestricted Customers |
| Class 2 | Domestic Economy 7 Customers |
| Class 3 | Non-Domestic Unrestricted Customers |
| Class 4 | Non-Domestic Economy 7 Customers |
| Class 5 | Non-Domestic Maximum Demand (MD) Customers with a Peak Load Factor (LF) of less than 20% |
| Class 6 | Non-Domestic Maximum Demand Customers with a Peak Load Factor between 20% and 30% |
| Class 7 | Non-Domestic Maximum Demand Customers with a Peak Load Factor between 30% and 40% |
| Class 8 | Non-Domestic Maximum Demand Customers with a Peak Load Factor over 40% |

Of these eight, only two refer to domestic properties although the profiles take into account the season and the day of the week. As an example of the standard profiles, Figure 1 shows the profiles for autumn for weekdays, Saturday and Sundays both for Economy 7 customers and non-Economy 7 customers plotted as 48 half hourly readings over the day.

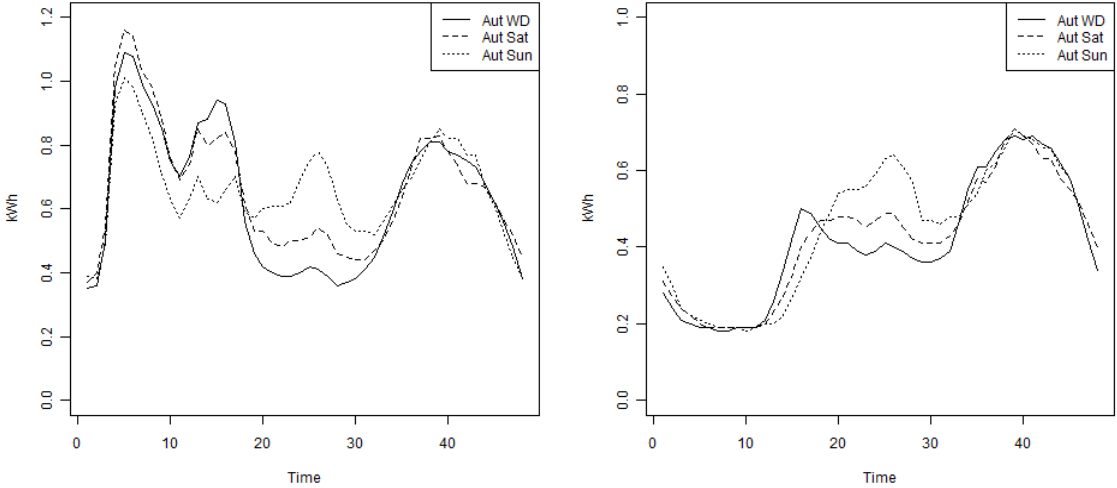

**Figure 1:** Economy 7 and non Economy 7 households in autumn

(Figueiredo et al., 2005) details the usage of Self organising maps and k-means clustering to define nine typical load profiles where the data is considered separately for weekdays and weekends and for each season. These clusters are then used as to classify, using C5.0 (Quinlan, 1993), data from the Portuguese electricity market consisting of 165 consumers in order to demonstrate the effectiveness of the clustering technique.

Various differing approaches have been taken to the definition of profiles in countries in Europe. As an example, in Croatia, (Marijanic & Karavidovic, 2007) detail a top down and a bottom up approach and after listing some of the issues conclude that a top down approach with each domestic customer having the same profile (scaled to total annual load) is the only practical short term solution.

The bottom up approach provides for a detailed build up of an overall profile and also provides insight into the reasons for the profile shape but does require a high level of monitoring within households so that each appliance and electrical circuit can be identified and aggregated. The detail available with this approach suggests it is preferable but the necessarily complex monitoring is likely to lead to a small sample population for analysis and hence over reliance on a few monitored households. It may not be possible to sample a sufficiently diverse population in order to gain a representative profile of the geographic area and hence aggregation may lead to errors.

The analysis of the overall electricity usage (as measured at the meter or at a single monitor) provides for a much simpler data collection method and is likely to lead to much higher numbers of households within the sample population due to the ease of collection. For this reason, the analysis of the overall electricity usage profile has been chosen as the focus of this work.

## 3 Methodology

This paper describes work in progress to apply the top-down framework approach detailed in (Figueiredo et al., 2005) to UK data and to extend and adapt the framework for UK specific conditions. The Figueiredo approach includes the following stages:

- Cleaning of the data in order to cope with missing data and outliers in the data.
- Normalisation of the data to make differing readings comparable.
- Splitting of the data into typical types of day such as weekday, weekend, holiday, hot day, etc. The Figueiredo work concentrates on weekend and weekday split and seasonal split only.
- Creation of representative daily load profiles. Various approaches can be taken at this stage and Figueiredo uses averaging across all available days within the day type and season.

- Application of a number of clustering techniques in order to group the data into a pre-defined number of clusters and then the definition of a representative load profile for each cluster.

The planned research on the UK data will review each of these steps and explore differing modifications that may be possible to improve the framework and its applicability to the UK electricity market.

The data used in this study is from an area of Milton Keynes, UK which was created in order to demonstrate various energy saving initiatives. The data was originally collected in 1988-91 by (Edwards, 1990) but was then stored on floppy disks which deteriorated physically and some of the original data has been lost. The original data disks were rescued and, where possible, regenerated by Steve Pretlove of UCL and, more recently, by Alex Summerfield with the work detailed in (Summerfield et al., 2007). The datasets have been made available in the UKERC data store.

The data is provided by the UKERC in the format of comma separated files with each dwelling having its own file of data. The energy data for the dwelling has been stored with some environmental data (e.g. rainfall, wind speed) for each hour that the energy data has been collected. All environmental data was loaded into a separate table as the data is dependent on the geographic location (i.e. Milton Keynes) rather than each individual house.

The original data was stored with a record for every hourly reading per property but it is more useful to store all the readings for a given day/property together in order that a profile of the usage over the day can be examined. MySQL was used to reshape the data with all the data for timed electricity meter readings being stored in a single table.

Some of the data readings are missing either due to the way in which the data was recovered from floppy disks or because of issues with the original collection of the data. For an initial view of the data, it was decided to omit all the days which contained a missing reading. However a method of replacing the missing values will be examined in future work based on the algorithm of:
- Calculate the average reading for a given property and given hour in the day from all the data with good readings.
- For each day with missing data, examine the data that is present and, by comparing with the overall average, calculate a measure of the day being considered as a fraction of the average day.
- Create missing data using the calculated fraction of the average day by multiplying the calculated fraction by the overall average for that hour of the day.

The extension of this method by calculating averages for different kinds of days (e.g. weekend and weekday, hot day) and replacing the data using the appropriate day type will also be researched.

Investigation is planned into various definitions of similarity between load profiles. The definition of "similar" is very dependent on the use being made of the results and various approaches will be investigated. For example, two households may have a similar shape of daily usage (e.g. high in morning, low during day, high late in the evening) but with very different total usage figures (i.e. the amplitude of the two household profiles varies). The shape of the pattern is often the key information as, for example, two consumers could have a similar usage pattern but one prefers a much warmer house and thus their actual energy readings are higher. If the analysis is to identify households using electricity at similar times across the day then the two households could be assessed as "similar". However, if the intention is to identify households using significantly more electricity than their neighbours then these two households should be seen as dissimilar.

The splitting of the data into differing "types" of day (e.g. weekend, holiday) will be extended by making use of the environmental data to allow differentiation based on wind speed (i.e. windy days versus calm days) or on temperature (i.e. hot versus cool days) rather on the blunter approach of splitting by season. With the variability of the UK weather this may provide more accurate clusters than splitting purely by season.

## 4  Initial Results

Initial investigation of the framework approach has been done using the Milton Keynes data which consists of 56601 fully populated rows and 7420 rows with errors representing data from 93 properties. Each row represents 24 hourly readings across the day for a given household.

Table 2 shows the variability in valid days of readings for the properties:

**Table 2:** Analysis of readings

|  | *Minimum number of readings per property* | *Maximum number of readings per property* | *Mean number of readings per property* |
|---|---|---|---|
| Valid readings | 199 | 700 | 608.3 |
| All readings | 281 | 819 | 688.4 |

This shows that numbers of readings vary from 199 days to 700 days - a large difference which means that averaging over all the properties should be done carefully. In a lot of households, there will be no data for particular dates and thus the averages for some dates may be based just on those properties with long collection periods and many valid readings.

The average hourly readings for each property were calculated over all valid data and then clustering (using k-means) was used in order to identify clusters of properties with a similar usage profile. The k-means clustering method relies on a random starting situation and requires the number of clusters as an input. In order to minimise the effects of the random starting point, the clustering algorithm was run 1000 times with differing random starting points and the best solutions taken.

Examination of the results shows that the large number of runs allows the same optimum solution to be found regardless of the initial random seed.

In order to find the best number of clusters to input to the algorithm, a series of clustering runs were done with cluster numbers from 2 to 10. The results of this can be seen at Figure 2.

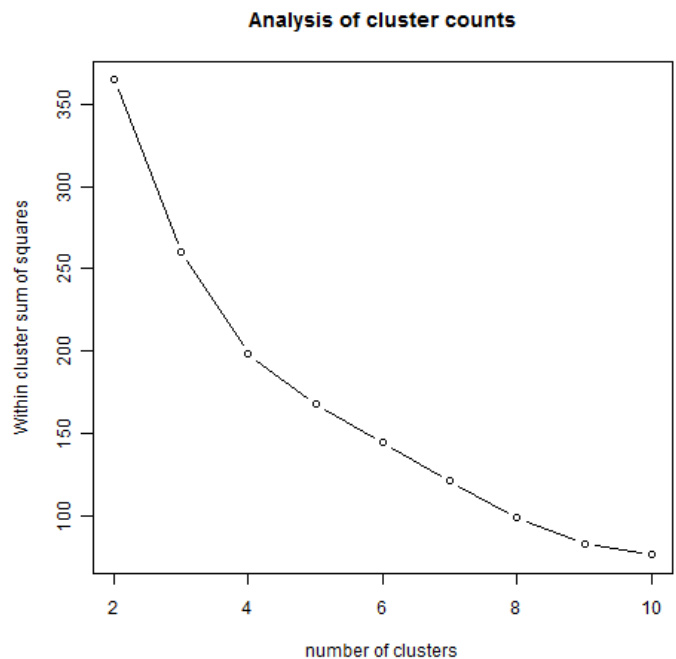

**Figure 2:** Analysis of Cluster Counts

The within cluster sum of squares was calculated for each of the input numbers of clusters. As the number of clusters increases, the total sum of squares will decrease (with the extreme example of each instance being in its own cluster with a total "within cluster sum of squares" being 0) and the graph at Figure 2 should be examined to find an obvious "elbow" that denotes an appropriate number of clusters. Whilst not strongly obvious, the elbow in the graph can be seen to be at number of clusters=4 and this was taken as the optimum number of clusters for further analysis.

The results of the clustering exercise on the Milton Keynes data are displayed at Figure 3 with the representative profile for the given cluster plotted in red and the average (over all valid days) profile for each of the households allocated to that cluster in black. The representative profile is that of the centroid calculated by the k-means clustering technique and is the profile that minimises the sum of squares between the centroid and the members of that cluster.

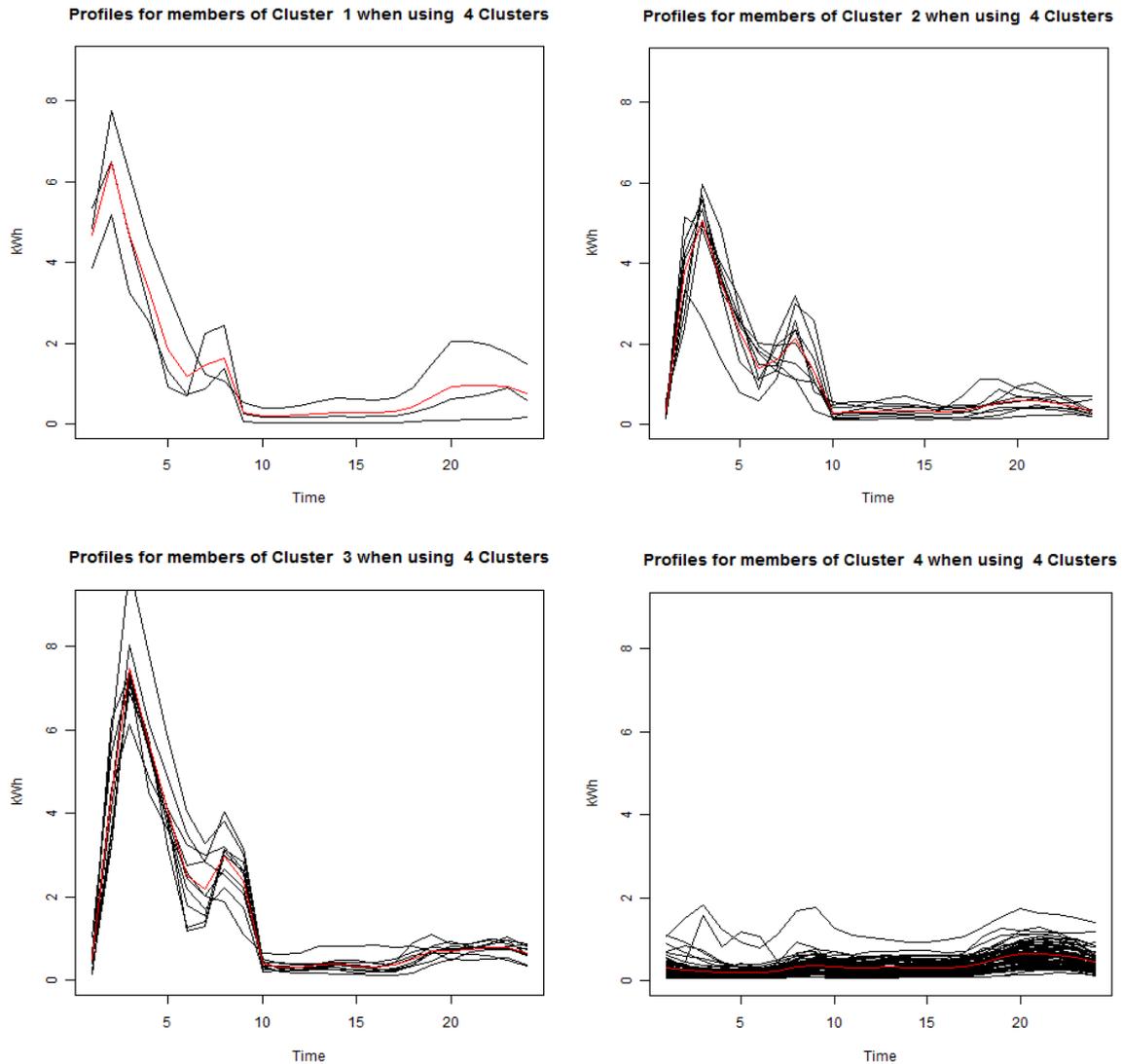

**Figure 3:** Profiles of members of each cluster

## 5 Analysis

The numbers of households in each cluster show a large number within cluster4 which appears to be the cluster of households not using Economy7. The other 3 clusters have much fewer households and reflect differing Economy7 users and the differing clusters may be indicative of using different appliances with high loads (for example, hot water heating with an immersion heater).

The graphs show that the clustering approach has led to grouping of households with similar shape profiles but with varying total amounts of electricity usage. For example, cluster1 shows similar shapes but a fairly wide spread of kWh readings.

The industry standard load profiles (see Figure 1) are significantly different from the typical load profiles identified but further investigation is needed in order to split the data by day type and season in order to allow more valid comparisons. The industry load profiles suggest a lower overall usage of electricity than that found in Milton Keynes in 1990 and may be a result of changes in domestic appliance efficiencies and how household behaviours have changed over time. In particular, storage heaters are much less popular than they were in the early 1990s.

# 6 Conclusions and Further Work

The differing approaches to building a household load profile have been considered with a "bottom-up" approach considered to be more explanatory of the makeup of the overall profile and providing more variability for "what-if" analyses but with a much higher effort requirement during data collection. For this reason, an approach focusing on the overall electricity usage of a household (as measured at a single meter and requiring no labelling of the usage by the householders or others) is preferred due to the much higher number of households that can be monitored.

A framework approach to determining clusters of similar electricity usage patterns has been laid out together with initial simple results of the clustering. Enhancement of the framework will be undertaken in future work with emphasis on missing values, the differentiation of recorded days into different types of day, and the differing definitions of similarity that could be used when assessing how to group households. Comparison of the clustering approaches and measurement of the suitability of each clustering algorithm will be undertaken.

The same approach will be applied on more recent UK data which is likely to have less usage of Economy7 (night time) electricity usage due to changing tariffs since 1990 and the much reduced usage of storage heaters.

The variable climate within the UK suggests that introducing temperature into the way in which the types of days are defined is worthy of investigation. Other countries may have very predictable daily temperatures (and hence heating usage) based on the seasons whilst the UK weather may require heating usage during the summer and also sometimes provide very warm days during winter. The differing ways of defining day types will be investigated and analysis done on how differing splits leads to differing or similar membership of the resulting clusters.

# 7 Acknowledgements

This data was accessed through the UK Energy Research Centre Energy Data Centre (UKERC-EDC). Our acknowledgments to the Building Research Establishment, which provided access to the original 1990 data set from Milton Keynes Energy Park, and to Bartlett School of Graduate Studies, University College London for processing and cleaning the raw data.

This work is possible thanks to EPSRC grant reference EP/I000496/1.

# 8 References

CAPASSO, A., GRATTIERI, W., LAMEDICA, R., & PRUDENZI, A. 1994. A bottom-up approach to residential load modeling. *Power Systems, IEEE Transactions on*, **9**(2), 957–964.
DECC. 2009. TOWARDS A SMARTER FUTURE: GOVERNMENT RESPONSE TO THE CONSULTATION ON ELECTRICITY AND GAS SMART METERING.

EDWARDS, J. 1990. Low energy dwellings in the Milton Keynes Energy Park. *Energy Management*, **26**, 32–33.

ELECTRICITY ASSOCIATION. 1997. Load profiles and their use in electricity settlement. *UKERC*.

FIGUEIREDO, V., RODRIGUES, F., VALE, Z., & GOUVEIA, J.B. 2005. An electric energy consumer characterization framework based on data mining techniques. *Power Systems, IEEE Transactions on*, **20**(2), 596–602.

MARIJANIC, T., & KARAVIDOVIC, D. 2007. Load profiling in an opening electricity market. *Pages 1–5 of: AFRICON 2007*. IEEE.

OFGEM. 2010. *Project Discovery - Options for delivering secure and sustainable energy supplies*.

PAATERO, J.V., & LUND, P.D. 2006. A model for generating household electricity load profiles. *International journal of energy research*, **30**(5), 273–290.

QUINLAN, J. ROSS. 1993. *C4.5: programs for machine learning*. San Francisco, CA, USA: Morgan Kaufmann Publishers Inc.

RAMOS, S., & VALE, Z. 2008. Data Mining techniques to support the classification of MV electricity customers. *Pages 1–7 of: Power and Energy Society General Meeting-Conversion and Delivery of Electrical Energy in the 21st Century, 2008 IEEE*. IEEE.

RICHARDSON, I., THOMSON, M., INFIELD, D., & CLIFFORD, C. 2010. Domestic electricity use: A high-resolution energy demand model. *Energy and Buildings*, **42**(10), 1878–1887.

SUMMERFIELD, AJ, LOWE, RJ, BRUHNS, HR, CAEIRO, JA, STEADMAN, JP, & ORESZCZYN, T. 2007. Milton Keynes Energy Park revisited: Changes in internal temperatures and energy usage. *Energy and Buildings*, **39**(7), 783–791.